4

# Spin Wave Scattering in Ferromagnetic Cross.

Alexander Kozhanov[1], Alexander Anferov[3], Ajey P. Jacob[2], S. James Allen[3]

[1]Department of Physics and Astronomy, Georgia State University, Atlanta, GA, 30303 USA
[2]Exploratory Research Device and Integration, GLOBALFOUNDRIES, Albany, NY 12203, USA
[3]University of California at Santa Barbara, Santa Barbara, CA 93106 USA

**Spin wave scattering in the right angle ferromagnetic cross was measured. Shape anisotropy defined magnetization ground states at zero biasing magnetic fields. Scattering of the spin waves in the center of ferromagnetic cross is strongly dependent on the amplitude and angle of the biasing magnetic field. Micromagnetic simulations indicate that low in-plane biasing magnetic fields rotate the magnetization of the cross center while the arms stay axially magnetized due to the shape anisotropy. We discuss effect of biasing magnetic fields on the spin wave scattering and approaches to an effective spin wave switch based on the fabricated structure.**

*Index Terms*— Ferromagnetic Cross, Scattering, Magnetostatic Spin Waves, Spin Wave Logic, Switch.

## I. INTRODUCTION

RECENT advances in technology explore alternative computational approaches as transistor performance approaches to its physical limits. Spintronics offers one of the possible directions for the development of logic elements [1]. Magnetic bipolar transistors [3], spin MOSFETs [3] and spin torque transfer devices [4-6] explore potential enhanced performance by sensing or using the spin degree of freedom that accompanies the charge current flow [7]. Spin waves can transfer spin information and have the potential for spin control without directly moving charge.

Spin waves in bulk materials have been studied for over a century. Bulk yttrium iron garnet (YIG) is used in a number of microwave applications such as delay lines, tunable filters, in which its ferromagnetic resonance is tuned by external magnetic field [8]. However, ease of deposition, processing and nanofabrication of ferromagnetic metals makes them more attractive for future nanoscale microwave devices. Further, ferromagnetic metals like CoTaZr and CoFeB have nearly an order of magnitude larger saturation magnetization than typical ferrimagnets [9]. As a result, they support higher and broader bands of spin wave resonances. Among the metallic ferromagnetic materials CoTaZr exhibits the lowest coercive fields of 2–10 Oe which enables using the shape anisotropy for self-alignment of the magnetization in the patterned ferromagnetic structures.

Shape anisotropy defines the magnetization alignment in the patterned ferromagnetic films. These structures support shape defined magnetostatic spin wave modes quantized by the structure dimensions at zero biasing magnetic field. Stripes and wires of various geometries, magnetic dots and antidot arrays, and tubes have been intensively studied. Quantized spin wave modes [10], [11], spin wave "tunneling" [12-15], current induced Doppler shifts [16], nonreciprocal spin wave propagation [17], magnon Bose-Einstein condensation [18] and various nonlinear effects [19], [20] have been observed in these structures.

Local magnetization inhomogenities artificially generated by local currents [13], transferred spin torque effects [21], [22], and magnetostriction [23] can be used to generate, detect and control the spin wave propagation. DC currents flowing in the wires placed in vicinity of the magnetic structure produce local magnetic field used to disturb the magnetization orientation in the part of the spin waveguide. This effect is used to manipulate the spin wave phase and amplitude [13]. A number of spin wave logic gates based on this effect were proposed [24], [25]. Most of the studied structures had the shape of either thin ferromagnetic film or ferromagnetic wire. However future development of spin wave based logic circuits relies on the interference devices which require studying structures with three and more spin wave wires joined in the junction device. Such structures require bends and change in the spin waveguide shape in order to provide the require spin wave propagation direction. Recent experiments on the right angle bends of the spin waveguide provided a nice demonstration of shape anisotropy importance for spin wave based logic devices [26].

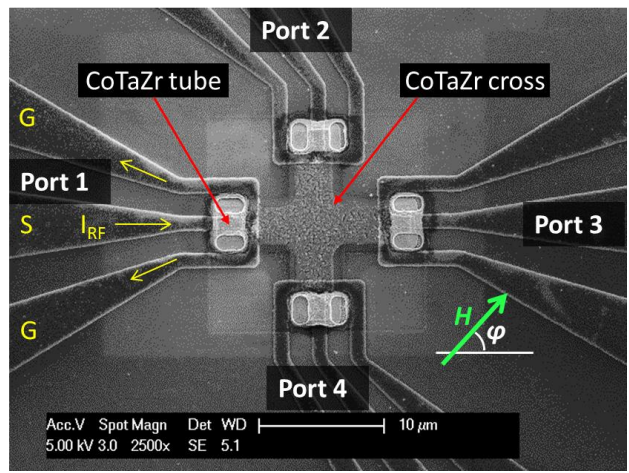

Fig. 1. Top view SEM micrograph of the CoTaZr cross with coupling loops and tube couplers at the cross arm ends. Magnetic field direction angle is measured from the horizontal (port 1 to port 3) line.

In this work we study backward volume spin wave propagation in a more complex magnetic structure which has a shape of spin waveguides crossing at a right angle, thus forming a ferromagnetic cross junction (Fig.1). We demonstrate that the spin wave scattering in the cross center is strongly dependent on the magnetization orientation of the cross center. Our experimental results indicate importance of spin wave scattering in ferromagnetic cross for spin wave logic and signal processing applications.

## II. Experimental approach

Ferromagnetic CoTaZr cross junction was lithographically fabricated using the following fabrication flow. A 200 nm thick amorphous ferromagnetic $Co_{90}Ta_5Zr_5$ film was sputtered onto a $Si/SiO_2$ substrate. A saturation magnetization of $M_s$=955 emu/cm3 and a coercive field $H_c$~2Oe were measured on the unpatterned $Co_{90}Ta_5Zr_5$ film using a vibrating sample magnetometer. More detailed information on the magnetic properties of sputtered $Co_{90}Ta_5Zr_5$ films used in our experiments can be found in the literature. Using lithographic techniques, the film was selectively etched in the ICP dry etch system with chlorine based chemistry to define the ferromagnetic cross with 4 μm wide and 12 μm long arms. The patterned ferromagnetic film was then covered with a 100 nm thick insulating $SiO_2$ layer. 100 nm thick aluminum coupling loops were lithographically formed by short-circuiting the ends of a pair of coplanar waveguides positioned over the cross arms. The structure was covered with another $SiO_2$ insulating layer, 200 nm thick. 2 μm long ferromagnetic tubes were formed by dry-etching holes down to the bottom magnetic layer and subsequently patterning another layer of $Co_{90}Ta_5Zr_5$ on top. Tubes thus formed at ends of the cross junction arms formed a closed magnetic circuit and served for enhanced coupling between microwaves and spin waves. More details about these tube couplers can be found in literature [27].

Transmission S-parameters were measured at room temperature using an Agilent 8720ES vector network analyzer (VNA) operating from 0.05 to 20 GHz. Test structure was probed using Cascade microtech infinity probes connected to the VNA ports. Only transmission $S_{21}$, $S_{31}$ and $S_{41}$ scattering parameters were measured (indexes correspond to the test structure port numbers). Spin waves were generated at the port 1 of the fabricated structure. Scattered waves were detected at ports 2 and 3. The test devices were positioned on the narrow gap of a small electromagnet that provided magnetic field bias up to 1000 Oe. By comparing the S-parameters at disparate bias magnetic fields, the magnetic field independent instrument response and electromagnetic coupling between measurement ports can be effectively removed to expose the S-parameters related to the magnetostatic mode coupling of exciting and detecting loops of the coplanar waveguides. Magnetic field orientation was varied in-plane of the test structure. The magnetic field orientation angle measured with respect to the line connecting ports 1 and 3 of the fabricated structure was varied within 0 ÷ 90 degrees range.

## III. Results and Discussion

In absence of external magnetic field we did not detect any transmission related to the magnetic coupling between exciting and detecting coupling loops at any of the 3 output ports due to the misalignment of the magnetization in the cross (Fig.1). As the magnetic field magnitude increased spin wave scattering into arms 2, 3 and 4 was detected starting at H~250 Oe. With increase of the magnetic field spin wave resonances shift towards higher frequencies and the transmission amplitude grows to saturation at $H$~700Oe.

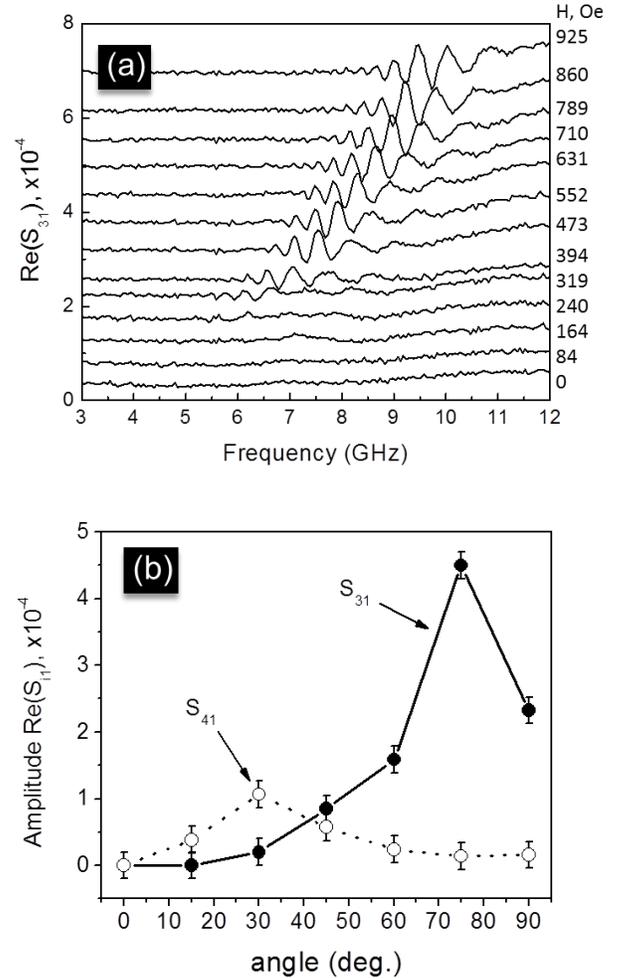

Fig.2. (a) Frequency and biasing magnetic field dependence of |S31| measured with the fabricated sample at $\varphi$=45°. (b) $S_{31}$ and $S_{41}$ dependence on the angle of external magnetic field measured at H=394 Oe.

Amplitude of the transmitted magnetostatic signal measured at ports 2-4 showed strong dependence on the angle of the external magnetic field. Almost no spin wave scattering into arms 2-4 was detected when the cross was biased with external magnetic field applied at $\varphi$=0° (measured transmission S parameters were in order of ~$10^{-5}$). The angular dependence of $S_{31}$ and $S_{41}$ real part amplitudes are shown in Fig.2b. $S_{21}$ and $S_{41}$ had identical angular dependences within the measurement error. Amplitude of the spin wave scattered into the opposite arm ($S_{31}$) reaches its maximum at $\varphi \approx 75^o$. $S_{21}$ and $S_{41}$ (spin waves scattered at the right angles into arms 2 and 4) had the maximum amplitude at φ≈30°. At φ≈45° spin waves scattered to the ports 2, 4 and port 3 has almost equal amplitudes.

Micromagnetic simulations show that at zero external magnetic fields ferromagnetic cross has four possible ground states with axially magnetized arms and the center magnetized at 45° with respect to the arms. However the ends of the arms experience magnetization distortions for the cross with short and wide arms. In the measured structure such magnetization distortions are even more enhanced by the tube couplers. Short tube couplers tend to be magnetized circularly along the tube perimeter [28]. Such magnetization layout is not favorable for spin wave excitation. This poor coupling is responsible for the absence of transmission measured at zero magnetic fields.

As the external magnetic field is increased the magnetization of the cross arms is "pinned" by the shape anisotropy while the magnetization of the cross center and the tube couplers is being manipulated. Magnetization orientation in the tube couplers and the cross center magnetization define the amplitude of the detected spin wave signal. However as can be seen from in Fig. 2b the strongest signal of the spin wave scattered to the opposite arm of the cross ($S_{31}$) is measured in the non-favorable orientation of the magnetization of the tube couplers, and almost no signal is measured at arms 2 and 4 for the most favorable orientation of the tube coupler's magnetization. This tells us that although there should be an effect of the spin wave excitation and detection efficiency on the measured scattered spin wave transmission signal, the magnetization orientation of the cross center significantly affects the spin wave scattering in this structure.

IV. CONCLUSIONS

In summary, spin wave scattering in the ferromagnetic cross was investigated using spin wave propagation spectroscopy. We have shown that spin wave scattering in this structure is strongly dependent on the magnetization orientation of the cross junction center. The tube couplers used for spin wave excitation and detection should affect the excited spin wave amplitude and coupling-out the spin wave signal; however, the experimental results indicate that distortions in the magnetization of the tube couplers do not define the measured transmission signal. However our experiments are not sufficient of providing quantitative analysis of the spin wave scattering. For more detailed analysis of the spin wave scattering processes cross junctions with locally controlled center magnetization and more controllable spin wave coupling-in and coupling-out should be explored.

The fabricated structure is potentially important for spin wave logic/signal processing applications. We demonstrated spin wave switching dependent on the local the magnetization of the ferromagnetic cross center. Magnetization of the cross junction center might be controlled by spin transfer torque device placed at the cross center. Input and output spin waveguides should have ferromagnetic wire geometry to provide sufficient shape anisotropy for magnetization alignment.

V. ACKNOWLEDGEMENTS

This work was supported by Nano Electronics Research Corporation (NERC) via the Nanoelectronics Research Initiative (NRI) at the Western Institute of Nanoelectronics (WIN) Center. The authors are also thankful to the staff of the UCSB ECE nanofabrication facilities for helpful discussions of the fabrication processes.

REFERENCES

[1] S.A. Wolf, D.D. Awschalom, R.A. Buhrman, J.M. Daughton, S. Von Molnar, M.L. Roukes, A.Yu Chtchelkanova, and D.M. Treger, "Spintronics: A spin-based electronics vision for the future," *Science* 294, 5546, pp.1488-1495, 2001.
[2] M. Flatté, M. E., Z. G. Yu, E. Johnston-Halperin, and D. D. Awschalom, "Theory of semiconductor magnetic bipolar transistors," *Appl.Phys.Lett.* 82, no. 26, pp.4740-4742, 2003.
[3] S. Datta, and B. Das, "Electronic analog of the electro-optic modulator," *Appl.Phys.Lett.* 56, pp.665-667, 1990.
[4] John C. Slonczewski, "Current-driven excitation of magnetic multilayers," *J. Magn. Magn. Mater.* 159, ppL1-L7, 1996.
[5] L. Berger, "Emission of spin waves by a magnetic multilayer traversed by a current," *Phys.Rev.B* 54, p.9353, 1996.
[6] E.B. Myers, D.C. Ralph, J.A. Katine, R.N. Louie, and R.A. Buhrman, "Current-induced switching of domains in magnetic multilayer devices," *Science* 285, pp.867-870, 1999.
[7] I. Žutić, J. Fabian, and S. Das Sarma, "Spintronics: Fundamentals and applications," *Rev.Mod.Phys.* 76, p.323, 2004.
[8] J.D. Adam, "Analog signal processing with microwave magnetics," *Proc.IEEE*, vol. 76, no. 2, pp. 159–170, Feb. 1988.
[9] B. Kuanr, I.R. Harward, D.L. Marvin, T. Fal, R.E. Camley, D.L. Mills, and Z. Celinski, "High-frequency signal processing using ferromagnetic metals," *IEEE Trans. Magn.*, 41, no. 10, pp.3538-3543, 2005.
[10] J. Jorzick, C. Kramer, S.O. Demokritov, B. Hillebrands, B. Bartenlian, C. Chappert, D. Decanini, "Spin wave quantization in laterally confined magnetic structures," *J.Appl.Phys.* 89, pp.7091-7095, 2001.
[11] A. Kozhanov, D. Ouellette, Z. Griffith, M. Rodwell, A.P. Jacob, D.W. Lee, S. X. Wang, and S. J. Allen, "Dispersion in magnetostatic CoTaZr spin waveguides," *Appl.Phys.Lett.* 94, pp.012505-012505, 2009.
[12] T. Schneider, A.A. Serga, A.V. Chumak, B. Hillebrands, R.L. Stamps, and M.P. Kostylev, "Spin-wave tunnelling through a mechanical gap," *Europhys.Lett.* 90, p.27003, 2010.
[13] S.O. Demokritov, A.A. Serga, A. Andre, V.E. Demidov, M.P. Kostylev, B. Hillebrands, and A.N. Slavin, "Tunneling of dipolar spin waves through a region of inhomogeneous magnetic field," *Phys.Rev.Lett.* 93, p.47201, 2004.
[14] U. Hansen, M. Gatzen, V.E. Demidov, and S.O. Demokritov, "Resonant tunneling of spin-wave packets via quantized states in potential wells," *Phys.Rev.Lett.* 99, p.127204, 2007.
[15] A. Kozhanov, D. Ouellette, M. Rodwell, S.J. Allen, A.P. Jacob, D.W. Lee, and S.X. Wang, "Dispersion and spin wave "tunneling" in nanostructured magnetostatic spin waveguides," *J.Appl.Phys.* 105, pp.07D311-07D311, 2009.
[16] V. Vlaminck, and M. Bailleul, "Current-induced spin-wave Doppler shift," *Science* 322, pp.410-413 (2008).
[17] P.K. Amiri, B. Rejaei, M. Vroubel, and Y. Zhuang, "Nonreciprocal spin wave spectroscopy of thin Ni–Fe stripes," *Appl.Phys.Lett.* 91, pp.062502-062502, 2007.
[18] S.O. Demokritov, V.E. Demidov, O. Dzyapko, G.A. Melkov, A.A. Serga, B. Hillebrands, and A.N. Slavin, "Bose–Einstein condensation of quasi-equilibrium magnons at room temperature under pumping," *Nature* 443, pp.430-433, 2006.
[19] V.E. Demidov, J. Jersch, K. Rott, P. Krzysteczko, G. Reiss, and S.O. Demokritov, "Nonlinear propagation of spin waves in microscopic magnetic stripes," *Phys.Rev.Lett.* 102, p.177207, 2009.
[20] V.E. Demidov, M. Buchmeier, K. Rott, P. Krzysteczko, J. Münchenberger, G. Reiss, and S.O. Demokritov, "Nonlinear

Hybridization of the Fundamental Eigenmodes of Microscopic Ferromagnetic Ellipses," *Phys.Rev.Lett.* 104, p.217203, 2010.

[21] V.E. Demidov, S. Urazhdin, and S.O. Demokritov, "Direct observation and mapping of spin waves emitted by spin-torque nano-oscillators," *Nature materials* 9, pp.984-988, 2010.

[22] C.T. Boone, J.A. Katine, J.R. Childress, V. Tiberkevich, A. Slavin, J. Zhu, X. Cheng, and I.N. Krivorotov, "Resonant nonlinear damping of quantized spin waves in ferromagnetic nanowires: a spin torque ferromagnetic resonance study," *Phys.Rev.Lett.* 103, p.167601, 2009.

[23] A. Khitun, D.E. Nikonov, and K.L. Wang, "Magnetoelectric spin wave amplifier for spin wave logic circuits," *J.Appl.Phys.* 106, pp.123909-123909, 2009.

[24] T. Schneider, A.A. Serga, B. Leven, B. Hillebrands, R.L. Stamps, and M.P. Kostylev, "Realization of spin-wave logic gates," *Appl.Phys.Lett.* 92, pp.022505-022505, 2008.

[25] A. Khitun, M. Bao, and K.L. Wang, "Magnonic logic circuits," *J.Phys.D* 43, p.264005, 2010.

[26] K. Vogt, H. Schultheiss, S. Jain, J.E. Pearson, A. Hoffmann, S.D. Bader, and B. Hillebrands, "Spin waves turning a corner," *Appl.Phys.Lett.* 101, pp.042410-042410, 2012.

[27] A. Kozhanov, D. Ouellette, M. Rodwell, S.J. Allen, D.W. Lee, S.X. Wang, "Micro-structured ferromagnetic tubes for spin wave excitation," *J.Appl.Phys.* 109, pp.07D333-1 – 07D333-3, 2011.

[28] A. Kozhanov, A., M. Popov, I. Zavislyak, D. Ouellette, D. W. Lee, S. X. Wang, M. Rodwell, and S. J. Allen. "Spin wave modes in ferromagnetic tubes." *J.Appl.Phys.* 111, pp.013905-013905, 2012.